\documentclass[pra,twocolumn,showpacs,letterpaper,showpacs,superscriptaddress]{revtex4}
\begin{document}
\newcommand{\bd}{{\bf d}}
\newcommand{\bv}{{\bf v}}
\newcommand{\hbp}{\hat{\bp}}
\newcommand{\hbx}{\hat{\bx}}
\newcommand{\hq}{\hat{q}}
\newcommand{\hp}{\hat{p}}
\newcommand{\ha}{\hat{a}}
\newcommand{\ad}{a^{\dag}}
\newcommand{\hsig}{{\hat{\sigma}}}
\newcommand{\nt}{\tilde{n}}
\newcommand{\itf}{\sl}
\newcommand{\eps}{\epsilon}
\newcommand{\bsig}{\pmb{$\sigma$}}
\newcommand{\beps}{\pmb{$\eps$}}
\newcommand{\bmu}{\pmb{$\mu$}}
\newcommand{\balpha}{\pmb{$\alpha$}}
\newcommand{\bbeta}{\pmb{$\beta$}}
\newcommand{\bgamma}{\pmb{$\gamma$}}
\newcommand{\bu}{{\bf u}}
\newcommand{\bpi}{\pmb{$\pi$}}
\newcommand{\bSig}{\pmb{$\Sigma$}}
\newcommand{\be}{\begin{equation}}
\newcommand{\ee}{\end{equation}}
\newcommand{\bea}{\begin{eqnarray}}
\newcommand{\eea}{\end{eqnarray}}
\newcommand{\sss}{_{{\bf k}\lambda}}
\newcommand{\ssss}{_{{\bf k}\lambda,s}}
\newcommand{\dip}{\langle\sigma(t)\rangle}
\newcommand{\dipp}{\langle\sigma^{\dag}(t)\rangle}
\newcommand{\sigz}{\langle\sigma_z(t)\rangle}
\newcommand{\sig}{{\sigma}}
\newcommand{\sigd}{{\sigma}^{\dag}}
\newcommand{\ra}{\rangle}
\newcommand{\la}{\langle}
\newcommand{\om}{\omega}
\newcommand{\Om}{\Omega}
\newcommand{\pa}{\partial}
\newcommand{\bR}{{\bf R}}
\newcommand{\bx}{{\bf x}}
\newcommand{\br}{{\bf r}}
\newcommand{\bE}{{\bf E}}
\newcommand{\bH}{{\bf H}}
\newcommand{\bB}{{\bf B}}
\newcommand{\bP}{{\bf P}}
\newcommand{\bD}{{\bf D}}
\newcommand{\bA}{{\bf A}}
\newcommand{\bek}{{\bf e}\rmk}
\newcommand{\rmk}{_{{\bf k}\lambda}}
\newcommand{\bsij}{{\bf s}_{ij}}
\newcommand{\bk}{{\bf k}}
\newcommand{\bp}{{\bf p}}
\newcommand{\epso}{{1\over 4\pi\eps_0}}
\newcommand{\BB}{{\mathcal B}}
\newcommand{\AAA}{{\mathcal A}}
\newcommand{\NN}{{\mathcal N}}
\newcommand{\mm}{{\mathcal M}}
\newcommand{\RR}{{\mathcal R}}
\newcommand{\bS}{{\bf S}}
\newcommand{\bL}{{\bf L}}
\newcommand{\bJ}{{\bf J}}
\newcommand{\bI}{{\bf I}}
\newcommand{\bF}{{\bf F}}
\newcommand{\bsub}{\begin{subequations}}
\newcommand{\esub}{\end{subequations}}
\newcommand{\baline}{\begin{eqalignno}}
\newcommand{\ealine}{\end{eqalignno}}
\newcommand{\isat}{{I_{\rm sat}}}
\newcommand{\Is}{I^{\rm sat}}
\newcommand{\Ip}{I^{(+)}}
\newcommand{\Imm}{I^{(-)}}
\newcommand{\Inu}{I_{\nu}}
\newcommand{\bInu}{\overline{I}_{\nu}}
\newcommand{\bN}{\overline{N}}
\newcommand{\qnu}{q_{\nu}}
\newcommand{\oqn}{\overline{q}_{\nu}}
\newcommand{\qsat}{q^{\rm sat}}
\newcommand{\Iout}{I_{\nu}^{\rm out}}
\newcommand{\topt}{t_{\rm opt}}
\newcommand{\crr}{{\mathcal{R}}}
\newcommand{\cE}{{\mathcal{E}}}
\newcommand{\cH}{{\mathcal{H}}}
\newcommand{\epsoo}{\epsilon_0}
\newcommand{\muo}{\mu_0}
\newcommand{\ombar}{\overline{\om}}
\newcommand{\bPi}{{\bf \Pi}}
\newcommand{\hz}{\hat{z}}

\title{Electromagnetic Momenta and Forces in Dispersive Dielectric Media}
\author{Douglas H. Bradshaw}
\affiliation{Los Alamos National Laboratory, Los Alamos, NM 87545 USA}
\author{Zhimin Shi}
\affiliation{The Institute of Optics, University of Rochester, Rochester, NY 14627 USA}
\author{Robert W. Boyd}
\affiliation{The Institute of Optics, University of Rochester, Rochester, NY 14627 USA}
\author{Peter W. Milonni}
\affiliation{104 Sierra Vista Dr, White Rock, NM 87544}

\date{\today}

\begin{abstract}
When the effects of dispersion are included, neither the Abraham nor the Minkowski expression for electromagnetic momentum  in a dielectric medium gives the correct recoil momentum for absorbers or emitters of radiation. The total
momentum density associated with a field in a dielectric medium has three contributions: (i) the Abraham momentum density of the field, (ii) the momentum density associated with the Abraham force, and (iii) a momentum density arising from the dispersive part of the response of the medium to the field, the 
latter having a form evidently first derived by D.F. Nelson [Phys. Rev. A{\bf 44}, 3985 (1991)]. All three 
contributions are required for momentum conservation in the recoil of an absorber or emitter in a dielectric
medium. We consider the momentum exchanged and the force on a polarizable particle (e.g., an
atom or a small dielectric sphere) in a host dielectric when a pulse of light is incident upon it, including the dispersion of the dielectric medium as well as a dispersive component in the response of the particle to the field. 
The force can be greatly increased in slow-light dielectric media. 
\end{abstract}


\maketitle
\section{Introduction}
Electromagnetic momentum in a dielectric medium is a subject with a very extensive literature, especially
in connection with its different formulations. The two most favored forms by far are those of Abraham
and Minkowski; as aptly remarked in a recent paper \cite{hinds}, ``There is ... a bewildering array of experimental studies and associated theoretical analyses which appear to favor one or other of these momenta or, indeed, others." 
An aspect of this subject that has received surprisingly little attention concerns the effects of dispersion on the
Minkowski and Abraham momenta and on the electromagnetic forces on polarizable particles. The intent of the present paper is to address such effects, which might help to clarify the physical interpretation of the Abraham and Minkowski momenta and the distinction between them. 

We first review briefly the Abraham and Minkowski momenta for the situation usually
considered---a dielectric medium assumed to be dispersionless and non-absorbing at a 
frequency $\om$. The Abraham and Minkowski momentum densities are respectively
\be
\bP_A = {1\over c^2}\bE\times\bH \ \ \ \ \ {\rm and} \ \ \ \ \ \bP_M=\bD\times\bB
\label{momA}
\ee
in the standard notation for the fields on the right-hand sides. We will take the 
permeability $\mu$ to be equal to its vacuum
value $\muo$, which is generally an excellent approximation at optical frequencies. For single
photons the magnitudes of the Abraham and Minkowski momenta are given by (see Section II)
\be
p_A={1\over n}{\hbar\om\over c} \ \ \ \ \ {\rm and} \ \ \ \ \ p_M=n{\hbar\om\over c}, 
\label{photA}
\ee
where $n$ is the refractive index at frequency $\om$. From $\bD=\epsoo n^2\bE$ it follows that
\be
{\pa\bP_M\over\pa t}={\pa\bP_A\over\pa t}+{\bf f}^A,
\label{momM}
\ee
where
\be
{\bf f}^A={1\over c^2}(n^2-1){\pa\over\pa t}(\bE\times\bH)
\label{abedensity}
\ee
is the {\sl Abraham force density}. For single-photon fields the momentum $p^A$ associated with the Abraham force
is $[(n^2-1)/n]\hbar\om/c$, and (\ref{momM}) becomes $p_M=p_A+p^A$.

The Abraham momentum is generally regarded as the correct momentum of the electromagnetic field \cite{jackson},
whereas the Minkowski momentum evidently includes the momentum of the dielectric medium as well as that of the field. Ginzburg \cite{ginz} calls $p_M$ the momentum of a 
``photon in a medium," and notes that
its use, together with energy and momentum conservation laws, yields correct results for Cerenkov radiation as
well as the Doppler shift. Experiments appear by and large to indicate that it is the momentum $n\hbar\om/c$
per photon that provides the recoil and radiation pressure experienced by an object immersed in a dielectric medium
\cite{jonesand}. However, when dispersion ($dn/d\om$) is accounted for, $n\hbar\om/c$ is not the Minkowski
momentum of a photon, as we review in the following section.

This paper is organized as follows. In the following section we briefly discuss the generalization of the Abraham and Minkowski momenta to the case of a dispersive dielectric medium \cite{garrison} and consider two examples: (i) the Doppler
shift in a dielectric medium \cite{fermi} and (ii) the displacement of a dielectric block on a frictionless surface due to the passage of a single-photon field through it \cite{nandor}. A consistent description of
momentum transfer in these examples requires that we account for momentum imparted to the medium. In Section  
\ref{forces} we calculate the force exerted by a quasimonochromatic plane wave on a polarizable particle
and on a dispersive dielectric medium modeled as a continuum, and obtain a dispersive contribution to the
latter in agreement with an expression that, to the best of our knowledge, was first derived, in a rather
different way, by Nelson \cite{nelson}. In Section \ref{momentumex} we consider the momentum exchange between
a plane-wave pulse and an electrically polarizable particle immersed in a nonabsorbing dielectric medium, and show that this momentum depends on both the dispersion of the medium and the variation with frequency of the polarizability; in particular, in slow-light media it can be large and in the direction opposite to that in which the field
propagates. Section \ref{sec:rayleigh} presents derivations of some results relevant to Section \ref{sec:sphere},
where we generalize the results of Section \ref{momentumex} to include absorption and discuss the forces 
exerted by a pulse on a small dielectric sphere in a host slow-light medium. Section \ref{sec:conclusions}
briefly summarizes our conclusions.

\section{Abraham and Minkowski Momenta for Dispersive Media}\label{abemink}
We first recall the expression for the total cycle-averaged energy density when a plane-wave monochromatic 
field [$\bE=\bE_{\om}e^{-i\om t}$, $\bH=\bH_{\om}e^{-i\om t}$, $\bH_{\om}^2=(\eps/\muo)\bE_{\om}^2$] propagates in a dispersive dielectric at a frequency $\om$ at which absorption is negligible \cite{landaulif}:
\be
u={1\over 4}\left[{d\over d\om}(\eps\om)\bE_{\om}^2+\muo\bH_{\om}^2\right],
\label{energydensity}
\ee
or equivalently, in terms of $\bE_{\om}$ and the group index $n_g=d(n\om)/d\om$,
\be
u={1\over 2}\epsoo nn_g\bE_{\om}^2.
\label{ueq}
\ee
When the field is quantized in a volume $V$, $u$ is in effect replaced by $q\hbar\om/V$, where $q$ is the expectation value of the photon number in the volume $V$;  therefore, from (\ref{ueq}), $\bE_{\om}^2$ is effectively $2\hbar\om/(\epsoo nn_gV)$ per photon. Thus, for single photons, the Abraham momentum defined by (\ref{momA}) is 
\be
p_A={n\over c}{1\over 2}\epsoo{2\hbar\om\over\epsoo nn_gV}V={1\over n_g}{\hbar\om\over c}.
\label{p1}
\ee
Similarly,
\be
p_M={n^2\over n_g}{\hbar\om\over c},
\label{p2}
\ee
which follows from the definition in (\ref{momA}) and the relation $\bD=\epsoo n^2\bE$;
thus $p_M=n^2p_A$. These same expressions for $p_A$ and $p_M$ can of course be obtained more formally by quantizing the fields $\bE$, $\bD$, $\bH$, and $\bB$ in a dispersive medium \cite{garrison}.

Two examples serve to clarify the differences among the momenta involved in the momentum exchange between light and matter. The first example
is based on an argument of Fermi's that the Doppler effect is a consequence of this momentum exchange \cite{fermi}, as 
follows. 
Consider an atom of mass $M$ inside a host dielectric medium with refractive index $n(\om)$. The atom has a sharply defined transition frequency $\om_0$ and is initially moving with velocity $v$ away from a source of light of frequency 
$\om$. Because the light in the atom's reference frame has a Doppler-shifted frequency $\om(1-nv/c)$ determined by the phase velocity ($c/n$) of light in the medium, the atom can absorb a photon
if $\om(1-nv/c)=\om_0$, or if
\be
\om\cong\om_0(1+nv/c). 
\label{dopp1}
\ee
We denote the momentum associated with a photon in the medium by $\wp$
and consider the implications of (nonrelativistic) energy and momentum conservation. The initial energy 
is $E_i=\hbar\om+{1\over 2}Mv^2$, and the final energy, after the atom has absorbed a photon, is
${1\over 2}Mv'^2+\hbar\om_0$, where $v'$ is the velocity of the atom after absorption. The initial momentum
is $\wp+Mv$, and the final momentum is just $Mv'$. Therefore
\be
{1\over 2}M(v'^2-v^2)\cong Mv(v'-v)=Mv(\wp/M)=\hbar(\om-\om_0),
\ee
or $\om\cong\om_0+\wp v/\hbar$. From (\ref{dopp1}) and $\om\cong\om_0$ we conclude that 
\be
\wp=n{\hbar\om\over c}.
\label{wp}
\ee
Thus, once we accept the fact that the Doppler shift depends on the refractive index of the medium according
to Eq. (\ref{dopp1}), we are led by energy and momentum conservation to conclude that an atom in the medium 
must recoil with momentum (\ref{wp}) when it absorbs (or emits) a photon of energy $\hbar\om$. Momentum
conservation in this example is discussed in more detail below.

In our second example we consider, following Balazs \cite{nandor}, a rigid block of mass $M$, refractive index $n$, and 
length $a$, initially sitting at rest on a frictionless surface. A single-photon pulse of frequency $\om$ passes through the block, which is assumed to be nonabsorbing at frequency $\om$ and to have anti-reflection coatings on its front and back surfaces. The length $a$ of the block is presumed to be much larger than the length of the pulse. If the photon momentum is $\wp_{\rm in}$ inside the block and $\wp_{\rm out}$ outside,
the block picks up a momentum $MV=\wp_{\rm out}-\wp_{\rm in}$ when the pulse enters. If the space outside the block is
vacuum, $\wp_{\rm out}=mc$, where $m=E/c^2=\hbar\om/c^2$. Similarly
$\wp_{\rm in}=mv_p$, where $v_p$ is the velocity of light in the block. Without dispersion,
$v_p=c/n$ and the momentum of the photon in the block is evidently $\wp_{\rm in}=mc/n=\hbar\om/nc$. 
The effect of dispersion is to replace $v_p=c/n$ by $v_g=c/n_g$ and $\wp_{\rm in}=\hbar\om/nc$ by $\wp_{\rm in}=\hbar\om/n_gc$. With or without dispersion, this example suggests that the photon momentum in the medium has the Abraham form. Note that the essential feature of Balazs's argument is simply that the velocity of light in the medium is $v_p$ (or, more generally, $v_g$). This, together with momentum conservation, is what leads him to conclude that
the momentum of the field has the Abraham form.

This prediction can in principle be tested experimentally. Conservation of momentum requires, according
to Balazs's argument, that $MV=m(c-v_g)$. When the pulse exits, the block recoils and comes to rest, and is left with a net displacement
\be
\Delta x=V\Delta t={m\over M}(c-v_g){a\over v_g}={\hbar\om\over Mc^2}(n_g-1)a
\ee
as a result of the light having passed through it. This is the prediction for the net displacement based on
the momentum $p_A$ given in (\ref{p1}). If the photon momentum inside the block were assumed to have the Minkowski form $n^2\hbar\om/cn_g$ given in (\ref{p2}), however, the displacement of the block would in similar fashion be predicted to be 
\be
\Delta x={\hbar\om\over Mc^2}a(n_g-n^2),
\ee
and if it were assumed to be $n\hbar\om/c$, as in Eq. (\ref{wp}), the prediction would be that the net
displacement of the block is
\be
\Delta x={\hbar\om\over Mc^2}an_g(1-n).
\ee
These different assumptions about the photon momentum can lead to different predictions not only for the
magnitude of the block displacement but also for its direction.

The first (Doppler) example suggests at first thought that the momentum of the photon is $n\hbar\om/c$ [Eq. (\ref{wp})], while the second (Balazs) example indicates that it is $\hbar\om/n_gc$. Let us consider more carefully the first example. There
is ample experimental evidence that the Doppler shift is $nv\om/c$ regardless of dispersion, 
as we have assumed, but does this imply that the momentum of a photon in a dielectric is in fact $n\hbar\om/c$?
We will show in the following section that the forces exerted by a plane monochromatic wave on the polarizable particles of a dielectric result in a momentum density of magnitude
\be
p_{\rm med}={\epsoo\over 2c}n(nn_g-1)E_{\om}^2=(n-{1\over n_g}){\hbar\om\over c}{1\over V};
\label{pmedium}
\ee
the second equality applies to a single photon, and follows from the replacement of $\bE_{\om}^2$ by 
$2\hbar\om/(\epsoo nn_gV)$, as discussed earlier. Now from the fact that
the Doppler shift implies that an absorber (or emitter) inside a dielectric 
recoils with momentum $n\hbar\om/c$, all
we can safely conclude from momentum conservation is that a momentum $n\hbar\om/c$ is taken from (or given to) 
the {\sl combined system of field
and dielectric}. Given that the medium has a momentum density (\ref{pmedium}) due to the force exerted on it by
the propagating field, we can attribute to the field (by conservation of momentum) a momentum density
\be
n{\hbar\om\over c}{1\over V}-P_{\rm med}={1\over n_g}{\hbar\om\over c}{1\over V}=p_A.
\ee
That is, the momentum of the field in this interpretation is given by the Abraham formula, consistent with the conclusion of the Balazs thought experiment. The recoil momentum $n\hbar\om/c$, which in general differs from both the Abraham and the Minkowski momenta, evidently gives the momentum not of the field as such but of the combined system of field plus dielectric. It is the momentum density equal to the {\sl total} energy density $u=\hbar\om/V$
for a monochromatic field divided by the phase velocity $c/n$ of the propagating wave. As already mentioned, experiments on the recoil of objects immersed in dielectric media have generally indicated that the recoil momentum is $n\hbar\om/c$ per unit of energy $\hbar\om$ of the field, just as in the Doppler effect. But this should not be taken to mean that $n\hbar\om/c$ is the momentum of a ``photon" existing independently of the medium in which the field propagates. Regardless of how this
momentum is apportioned between the field and the medium in which it propagates, the important thing for the theory,
of course, is that it correctly predicts the {\sl observable forces} exerted by electromagnetic fields. We next turn our attention specifically to the forces acting on polarizable particles in applied electromagnetic fields.

\section{Momenta and Forces on Polarizable Particles}\label{forces}
We will make the electric dipole approximation and consider field frequencies
such that absorption is negligible. Then the induced electric dipole moment of a particle in a field of frequency $\om$ is $\bd=\alpha(\om)\bE_{\om}\exp(-i\om t)$, and the polarizability $\alpha(\om)$
may be taken to be real. With these assumptions we now consider the forces acting on such particles in applied,
quasi-monochromatic fields. 

We begin with the Lorentz force on an electric dipole moment $\bd$ in an electromagnetic field \cite{barloud}:
\bea
\bF&=&(\bd\cdot\nabla)\bE+\dot{\bd}\times\bB\nonumber \\
&=&(\bd\cdot\nabla)\bE+\bd\times(\nabla\times\bE)+{\pa\over\pa t}(\bd\times\bB)\nonumber \\
&\equiv&\bF_E+\bF_B,
\label{dipforce}
\eea
where we define
\be
\bF_E=(\bd\cdot\nabla)\bE+\bd\times(\nabla\times\bE),
\label{forceE}
\ee
\be
\bF_B={\pa\over\pa t}(\bd\times\bB).
\label{forceB}
\ee
In writing the second equality in (\ref{dipforce}) we have used the Maxwell equation $\pa\bB/\pa t=-\nabla\times\bE$.
The dipole moment of interest here is induced by the electric field. Writing
\be
\bE=\cE_0(\br,t)e^{-i\om t}=e^{-i\om t}\int_{-\infty}^{\infty}d\Delta\tilde{\cE}_0(\br,\Delta)e^{-i\Delta t},
\ee
in which $|\pa\cE_0/\pa t|\ll\om|\cE_0|$ for a quasi-monochromatic field, we approximate $\bd$ as follows:
\bea
\bd(\br,t)&=&\int_{-\infty}^{\infty}d\Delta\alpha(\om+\Delta)\tilde{\cE}_0(\br,\Delta)e^{-i(\om+\Delta)t}\nonumber \\
&\cong& \int_{-\infty}^{\infty}d\Delta[\alpha(\om)+\Delta\alpha'(\om)]\tilde{\cE}_0(\br,\Delta)e^{-i(\om+\Delta)t}
\nonumber \\
&=&\left[\alpha(\om)\cE_0(\br,t)+i\alpha'(\om){\pa\cE_0\over\pa t}\right]e^{-i\om t}.
\label{eqd}
\eea
Here $\alpha'=d\alpha/d\om$ and we assume that higher-order dispersion is sufficiently weak that 
terms $d^m\alpha/d\om^m$ can be neglected for $m\ge 2$. Putting (\ref{eqd}) into (\ref{forceE}), we obtain 
after some straightforward manipulations and cycle-averaging the force
\be
\bF_E=\nabla\left[{1\over 4}\alpha(\om)|\cE|^2\right]+{1\over 4}\alpha'(\om)\bk{\pa\over\pa t}|\cE|^2,
\label{force11}
\ee
where $\cE$ and $\bk$ are defined by writing $\cE_0(\br,t)=\cE(\br,t)e^{i\bk\cdot\br}$. Since the refractive index $n$ of a medium in which local field corrections are negligible is given in terms of $\alpha$ by $n^2-1=N\alpha/\epsoo$, $N$ being the density
of dipoles in the dielectric, we have $\alpha'=(2n\epsoo/N)(dn/d\om)$ and
\be
\bF_E=\nabla\left[{1\over 4}\alpha(\om)|\cE|^2\right]+{\epsoo\over 2N}\bk n{dn\over d\om}{\pa\over\pa t}|\cE|^2.
\label{force1}
\ee
The first term is the ``dipole force" associated with the energy $W=-{1\over 2}\alpha(\om)\bE^2$ involved in inducing
an electric dipole moment in an electric field:
\be
W=-\int_0^{\bE}\bd\cdot d\bE=-\alpha(\om)\int_0^{\bE}\bE\cdot d\bE= -{1\over 2}\alpha(\om)\bE^2.
\ee
The second term in (\ref{force1}) is nonvanishing only because of dispersion
($dn/d\om\ne 0$). It is in the direction of propagation of the field, and implies for a uniform density $N$ of
atoms per unit volume a momentum density of magnitude
\be
P_D={1\over 2}\epsoo n^2{dn\over d\om}{\om\over c}|\cE|^2={1\over 2}{\epsoo\over c}n^2(n_g-n)|\cE|^2,
\label{nels}
\ee
since $k=n(\om)\om/c$. This momentum density comes specifically from the dispersion ($dn/d\om$) of
the medium.

The force $\bF_B$ defined by (\ref{forceB}), similarly, implies a momentum density $\bP^A$
imparted to the medium:
\be
\bP^A=N\bd\times\bB.
\ee
As the notation suggests, this momentum density is associated with the Abraham force density (\ref{abedensity}).
The result of a straightforward evaluation of $\bP^A$ based on (\ref{eqd}) and $\nabla\times\bE=-\pa\bB/\pa t$ is 
\be
\bP^A={1\over 2}\epsoo(n^2-1){\bk\over\om}|\cE|^2, \ \ \ \ P^A={1\over 2}{\epsoo\over c}n(n^2-1)|\cE|^2,
\label{fm}
\ee
when we use $\bk\cdot\bE=0$ and our assumption that $|\dot{\cE}_0|\ll\om|\cE_0|$. The magnitude of the total momentum density in the medium due to the force of the field on the dipoles is therefore
\bea
P_{\rm med}=P_D+P^A&=&{\epsoo\over 2c}\left[n^2(n_g-n)+n(n^2-1)\right]|\cE|^2\nonumber \\
&=&{\epsoo\over 2c}n(nn_g-1)|\cE|^2
\label{pmed}
\eea
in the approximation in which the field is sufficiently uniform that we can ignore the dipole force $\nabla[{1\over 4}\alpha|\cE|^2]$. 

The complete momentum density for the field and the medium is obtained by adding to (\ref{pmed}) the Abraham momentum 
density $P_A$ of the field. According to (\ref{momA}), $P_A=(\epsoo/ 2c)n|\cE|^2$, and so the total momentum density is
\be
P_A+P_D+P^A={\epsoo\over 2c}[n+n(nn_g-1)]|\cE|^2={\epsoo\over 2c}n^2n_g|\cE|^2
\label{tot}
\ee
if the dipole force is negligible.
To express these results in terms of single photons, we again replace $|\cE_0|^2$ by $2\hbar\om/(\epsoo nn_gV)$; 
then (\ref{tot}) takes the form
\be
p_A+p_D+p^A=n{\hbar\om\over c}{1\over V},
\label{ppmed}
\ee
consistent with the discussion in the preceding section. This is the total momentum density per photon, assuming that the dipole force is negligible. The momentum density of the medium per photon follows from (\ref{pmed}):
\bea
p_{\rm med}&=&p_D+p^A={\epsoo\over 2c}n(nn_g-1){2\hbar\om\over nn_g\epsoo V}\nonumber \\
&=&(n-{1\over n_g}){\hbar\om\over c}{1\over V},
\eea
as stated earlier [Eq. (\ref{pmedium})]. 

Consider the example of spontaneous emission by a guest atom in a host dielectric 
medium. The atom loses energy $\hbar\om_0$, and the quantum (photon in the
medium) of excitation carries away from the atom not only this energy but also a linear momentum 
$n\hbar\om/c$ [Eq. (\ref{ppmed})]. The atom therefore recoils with momentum $n\hbar\om/c$ \cite{pwmboyd}.

The momentum density (\ref{nels}) was obtained by Nelson \cite{nelson} in a rigorous treatment of a deformable
dielectric based on a Lagrangian formulation; in the present paper a dielectric medium is
treated as an idealized rigid body. From a microscopic perspective, this part of the momentum density
of the medium is attributable directly to the second term on the right-hand side of (\ref{eqd}), i.e., to the
part of the induced dipole moment that arises from dispersion. In the Appendix the relation of this term to 
the formula (\ref{energydensity}) for the total energy density is reviewed; the term is obviously a
general property of induced dipole moments in applied fields. Consider, for example, a two-level atom driven by
a quasi-monochromatic field with frequency $\om$ far-detuned from the atom's resonance frequency $\om_0$.
In the standard $u,v$ notation for the off-diagonal components of the density matrix in the rotating-wave
approximation \cite{jhe},
\be
u(t)-iv(t)\cong {1\over\Delta}\chi(t)+{i\over\Delta^2}{\pa\chi\over\pa t} + ... \ ,
\label{tla}
\ee
where $\chi(t)$ is the Rabi frequency and $\Delta$ is the detuning. The polarizability is proportional
to $1/\Delta$ in this approximation, and therefore (\ref{tla}) is just a special case of (\ref{eqd}).
 
\section{Momentum Exchange between a Light Pulse and an Induced Dipole}\label{momentumex}
We next consider the momentum exchange between a plane-wave {\sl pulse} and a single polarizable particle. We will assume again that the particle is characterized by a real polarizability $\alpha(\om)$ and that it is surrounded by a host medium with refractive index $n_b(\om)$. The electric field is assumed to be
\be
\bE(z,t)=\cE(t-z/v_{bg})\cos(\om t-kz),
\label{pulse}
\ee
with $k=n_b(\om)\om/c$ and group velocity $v_{bg}=c/n_{bg}$, $n_{bg}=(d/d\om)(\om n_{b})$.

The force acting on the particle is $\bF_E+\bF_B$. $\bF_B$ reduces to 
${1\over 2}\alpha(\om)(\bk/\om)(\pa/\pa t)|\cE|^2$, obtained by multiplying (\ref{fm}) by a volume $V$ 
describing the pulse, replacing $n^2-1$ by $N\alpha/\epsoo$ with $NV=1$ for the single particle, and
differentiation with respect to time. $\bF_E$ follows from (\ref{force11}). Then the force acting on the particle is in the $z$ direction and has the (cycle-averaged) magnitude 
\bea
F&=&{1\over 4}\alpha(\om){\pa\over\pa z}\cE^2+{1\over 4}\alpha'(\om)n_b(\om){\om\over c}{\pa\over\pa t}\cE^2\nonumber \\
&&\mbox{}+{1\over 2c}\alpha(\om)n_b(\om){\pa\over\pa t}\cE^2,
\label{forcey}
\eea 
where now we retain the dipole force, given by the first term on the right-hand side.
The momentum of the particle at $z$ at time $T$ is
\bea
p&=&\int_{-\infty}^T Fdt={1\over 4}\alpha\int_{-\infty}^T{\pa\over\pa z}\cE^2(t-z/v_{bg})dt\nonumber \\
&&\mbox+{1\over 4c}\alpha'n_b\om
\int_{-\infty}^T{\pa\over\pa t}\cE^2(t-z/v_{bg})dt\nonumber \\
&&\mbox{}+{1\over 2c}\alpha n_b\int_{-\infty}^T{\pa\over\pa t}\cE^2(t-z/v_{bg})dt\nonumber \\
&=&-{1\over 4}\alpha{1\over v_{bg}}\cE^2+{n_b\over 4c}\alpha'\om\cE^2+{1\over 2}\alpha{n_b\over c}\cE^2\nonumber \\
&=&{1\over 4c}[(2n_b-n_{bg})\alpha+n_b\om\alpha']\cE^2(T-z/v_{bg}).
\label{for1}
\eea

Hinds and Barnett \cite{hinds} have considered the force on a two-level atom due to a pulse of light in free space.
In this case $n_b=n_{bg}=1$ and (\ref{for1}) reduces to
\be
p={1\over 4c}[\alpha+\om\alpha']\cE^2.
\ee
Following Hinds and Barnett, we argue that a pulse occupying the volume $V$ in the neighborhood of the atom
in free space corresponds to a number $q={1\over 2}\epsoo\cE^2V/\hbar\om$ of photons, so that
\be
p={1\over 2c}[\alpha+\om\alpha']{\hbar\om\over\epsoo V}q.
\ee
$\alpha=\epsoo(n^2-1)/N$, where $n$ is the refractive index in the case of $N$ polarizable particles per unit 
volume. Then
\bea
p&=&{1\over 2c}\left[{\epsoo(n^2-1)\over N}+{2\epsoo n\over N}\om{dn\over d\om}\right]{\hbar\om\over c}q\nonumber \\
&\cong&[n-1+\om{dn\over d\om}]{\hbar\om\over c}q\equiv K{\hbar\om\over c}q.
\label{for2}
\eea
This is the momentum imparted to the particle, which implies a change in {\sl field} momentum per photon equal to 
\be
{\hbar\om\over c}[1-K]\cong{\hbar\om\over c}{1\over 1+K}={\hbar\om\over n_gc}
\ee
if $|K|\ll 1$, where $n_g=(d/d\om)(n\om)$. As in the case of a two-level atom considered by Hinds and Barnett,
this corresponds to the Abraham momentum; our result simply generalizes theirs in replacing $n$ by $n_g$
in the expression for the change in photon momentum.

In the case of a polarizable particle in a host dielectric rather than in free space we obtain, from 
(\ref{for1}),
\be
p={I\over 2\epsoo c^2}[(2-{n_{bg}\over n_b})\alpha+\om\alpha'],
\label{forr2}
\ee
where the intensity $I=(1/2)c\epsoo n_b\cE^2$. If dispersion in the medium and in the polarizability of the guest particle are negligible, we can set $n_{bg}=n$ and $\alpha'=0$, and then (\ref{forr2}) reduces to 
a well known expression \cite{gordon}. However, this momentum can be large in a slow-light 
medium ($n_{bg}$ large), for example, because the gradient of the field (\ref{pulse}) responsible 
for the dipole force on the particle is large \cite{slow}; this is a consequence of the spatial compression of
a pulse in a slow-light medium. We discuss this case further in Section \ref{sec:sphere}.
But first we return to some other well known results that are relevant there.

\section{Electric Dipole Radiation Rate and Rayleigh Scattering \cite{wod}}\label{sec:rayleigh}
A Hertz vector $\bPi(\br,\om)$ can be defined for a dielectric
medium, analogous to the case of free space \cite{bornwolf}, by writing the electric and magnetic field 
components at frequency $\om$ as
\be
\bE(\br,\om)=k_0^2[\eps_b(\om)/\epsoo]\bPi(\br,\om)+\nabla[\nabla\cdot\bPi(\br,\om)],
\label{her1}
\ee
\be
\bH(\br,\om)=-i\om\eps_b(\om)\nabla\times\bPi(\br,\om).
\label{her2}
\ee
Here $k_0=\om/c$ and we denote by $\eps_b(\om)$ the (real) permittivity of the dielectric. We will be interested here
in a dipole source inside the ``background" dielectric medium. The identifications (\ref{her1}) and (\ref{her2}) are consistent with the propagation of a wave of frequency $\om$ with the phase velocity $c/n_b(\om)$ in the 
medium [$n_b(\om)=\sqrt{\eps_b(\om)/\epsoo}$], as will be
clear in the following.

The curl of $\bE(\br,\om)$ in (\ref{her1}) is simply
\be
\nabla\times\bE(\br,\om)=k_0^2[\eps_b(\om)/\epsoo]\nabla\times\bPi(\br,\om),
\ee
since the curl of a gradient is zero. Now apply the curl operation to this equation, assuming no free currents and therefore $\nabla\times\bH(\br,\om)=-i\om\bD(\br,\om)$:
\bea
\nabla\times(\nabla\times\bE)&=&i\om\muo\nabla\times\bH=\om^2\muo\bD\nonumber \\
&=&k_0^2[\eps_b(\om)/\epsoo]\nabla\times(\nabla\times\bPi)\nonumber \\
&=&k_0^2[\eps_b(\om)/\epsoo][\nabla(\nabla\cdot\bPi)-\nabla^2\bPi],
\eea
implying
\be
\nabla^2\bPi={\epsoo\over\eps_b}{\om^2\over k_0^2}\muo\bD+\nabla(\nabla\cdot\bPi)=
-{1\over\eps_b}\bD+[\bE-{\eps_b\over\epsoo}k_0^2\bPi],
\ee
\be
\nabla^2\bPi+k^2\bPi=\bE-{1\over\eps_b}\bD, \ \ \ \ \ k^2=k_0^2\eps_b(\om)/\epsoo=n_b^2(\om)\om^2/c^2.
\ee
If $\bD(\br,\om)=\eps_b(\om)\bE(\br,\om)$, the right-hand side is zero, and all we have done is rederived
what we already know: the field propagates with phase velocity $\om/k(\om)=c/n_b(\om)$. Suppose, however,
that within the medium there is a localized source characterized by a dipole moment density $\bP_s(\br,\om)=
\bp_0(\om)\delta^3(\br)$. Then $\bD=\eps_b\bE+\bP_s$ and
\be
\nabla^2\bPi+k^2\bPi=-{1\over\eps_b}\bp_0(\om)\delta^3(\br).
\ee
The solution of this equation for $\bPi(\br,\om)$ is simply 
\be
\bPi(\br,\om)={1\over 4\pi\eps_b(\om)}\bp_0(\om){e^{ikr}\over r},
\label{her102}
\ee
and from this one obtains the electric and magnetic fields due to the source in the medium. In the far field,
assuming $\bp_0=p\hz$ and letting $\theta$ be the angle between the $z$ axis and the observation
point,
\be
E_{\theta}={k_0^2p\over 4\pi\epsoo}\sin\theta{e^{ikr}\over r},
\ee
\be
H_{\phi}={n_bk_0^2p\over 4\pi\epsoo}\sqrt{\epsoo\over\muo}\sin\theta{e^{ikr}\over r},
\ee
in spherical coordinates. The Poynting vector $\bS=\bE\times\bH$ implies the radiation rate
\be
P={n_bp^2\om^4\over 12\pi\epsoo c^3},
\label{here}
\ee
analogous to the fact that the spontaneous emission rate of an atom in a dielectric without local
field corrections is proportional to the (real) refractive index at the emission frequency.

\subsection*{Polarizability of a Dielectric Sphere}
Suppose, somewhat more generally, that the source within the medium occupies a volume $V$ and is characterized
by a permittivity $\eps_s(\om)$. Then $\bD(\br,\om)=\eps(\br,\om)\bE(\br,\om)$, where $\eps=\eps_s(\om)$ within the
volume $V$ occupied by the source and $\eps(\br,\om)=\eps_b(\om)$ outside this volume, and
\be
\nabla^2\bPi+k^2\bPi=[1-\eps(\br,\om)/\eps_b(\om)]\bE.
\ee
The solution of this equation is 
\be
\bPi(\br,\om)=-{1\over 4\pi}\left[1-{\eps_s(\om)\over\eps_b(\om)}\right]\int_V d^3r'\bE(\br',\om){e^{ik|\br-\br'|}\over|\br-\br'|}.
\nonumber \\
\label{her3}
\ee
Suppose further that the extent of the volume $V$ is sufficiently small compared to a 
wavelength that we can approximate (\ref{her3}) by
\be
\bPi(\br,\om)=-{1\over 4\pi}\left[1-{\eps_s(\om)\over\eps_b(\om)}\right]V\bE_{\rm ins}(\om){e^{ikr}\over r},
\ee
with $r$ the distance from the center of the source (at $\br=0$) to the observation point and $\bE_{\rm ins}(\om)$
the (approximately constant) electric field in the source volume $V$. This has the same form as (\ref{her102})
with $\bp_0(\om)=\eps_b(\om)[\eps_s(\om)/\eps_b(\om)-1]V\bE_{\rm ins}(\om)$. In other words, $\bPi(\br,\om)$
has the same form as the Hertz vector for an electric dipole moment
\be
\bp_0(\om)=[\eps_s(\om)-\eps_b(\om)]V\bE_{\rm ins}(\om).
\label{her4}
\ee
Consider, for example, a small dielectric sphere of radius $a$: $V=4\pi a^3/3$. The field inside such a sphere
is $\bE_{\rm ins}(\om)=[3\eps_b/(\eps_s+2\eps_b)]\bE_b(\om)$, where $\bE_b(\om)$ is the (uniform) electric field
in the medium in the absence of the source. The dipole moment (\ref{her4}) in this case is therefore related to the
external field $\bE_{\rm out}(\om)$ by $\bp_0(\om)=\alpha(\om)\bE_{\rm out}(\om)$, where the polarizability
\be
\alpha(\om)=4\pi\eps_b\left({\eps_s-\eps_b\over\eps_s+2\eps_b}\right)a^3.
\label{her5}
\ee

\subsection*{Rayleigh Attenuation Coefficient}
The cross section for Rayleigh scattering for an ideal gas of refractive index $n(\om)$ can be deduced as 
follows \cite{rayleighnote}. An electric field $\bE_0\cos\om t$ induces an electric dipole moment $\bp(t)=\alpha(\om)\bE_0\cos\om t$ in each of $N$ isotropic, polarizable particles
per unit volume, each particle having a spatial extent small compared to a wavelength. The power radiated by this 
dipole is, from Eq. (\ref{here}),
\be
{dW_{\rm rad}\over dt}=n(\om){\om^4\over 12\pi\epsoo c^3}\alpha^2(\om)\bE_0^2\equiv\sigma_{\rm R}(\om)I,
\label{rayleigh1}
\ee
where $W_{\rm rad}$ denotes energy of the radiated field, $I={1\over 2}n(\om)c\epsoo\bE_0^2$ is the intensity of the
field incident on the dipole, and 
\be
\sigma_{\rm R}(\om)={1\over 6\pi N^2}\left({\om\over c}\right)^4[n^2(\om)-1]^2
\label{rayleigh2}
\ee
is the (Rayleigh) scattering cross section. We have assumed that local field corrections are negligible and used the formula $n^2(\om)-1=N\alpha(\om)/\epsoo$ to express $\sigma_{\rm R}(\om)$ in terms of the refractive index $n(\om)$. The attenuation coefficient is then
\be
a_R=N\sigma_{\rm R}={1\over 6\pi N}\left({\om\over c}\right)^4[n^2(\om)-1]^2.
\label{ray1899}
\ee
Rosenfeld \cite{rosenfeld} obtains instead
\be
a_R=N\sigma_{\rm R}={1\over 6\pi n(\om)N}\left({\om\over c}\right)^4[n^2(\om)-1]^2,
\label{rosenfeld}
\ee
because he does not account for the factor $n(\om)$ in the dipole radiation rate (\ref{rayleigh1}). 
Rayleigh's derivation of
(\ref{ray1899}) follows essentially the one just given, but the factor $n(\om)$ appears in neither the dipole radiation
rate nor the expression for the intensity (or actually, in his derivation, the energy density) \cite{rayleigh}. In practice the difference
between (\ref{ray1899}) and (\ref{rosenfeld}) is negligible for the case assumed here of a dilute medium \cite{meas}.

\section{Force on a Dielectric Sphere}\label{sec:sphere}
The expression (\ref{forcey}) for the force on a polarizable particle in a field (\ref{pulse}) may be generalized
to allow for absorption by the particle simply by taking the polarizability $\alpha(\om)$ in (\ref{eqd}) to
be complex. Assuming again that $\cE$ is slowly varying in time compared to $\exp(-i\om t)$, and slowly varying in
space compared to $\exp(ikz)$, we obtain
\be
F={1\over 4c}[(2n_b-n_{bg})\alpha_R+n_b\om \alpha_R']{\pa\over\pa\tau}|\cE|^2
+{1\over 2}n_b{\om\over c}\alpha_I|\cE|^2,
\label{forceyy}
\ee
where $\tau=t-n_{bg}z/c$ and $\alpha_R$ and $\alpha_I$ are the real and imaginary parts, respectively, of $\alpha(\om)$. If we replace $n_{bg}$ by $n_b$ and take $\alpha_R'\cong 0$, we recover
results that may be found in many previous works when absorption is assumed to be negligible \cite{gordon}. The last term 
in (\ref{forceyy}) is the absorptive contribution to equation (7) of a paper by Chaumet and Nieto-Vesperinas 
\cite{chaumet} when the field is assumed to have the form (\ref{pulse}). 

The polarizability in the case of a dielectric sphere of radius $a$ much smaller than the wavelength of the field
is given by (\ref{her5}). Dispersion affects the force (\ref{forceyy}) both through the group index
($n_{bg}$) of the host dielectric medium and the variation of the real part of the sphere's polarizability with frequency ($\alpha_R'$). The latter
depends on both the intrinsic frequency dependence of the permittivity of the material of the sphere and the frequency dependence of the refractive index of the host medium. If these dispersive contributions to the force exceed the remaining two contributions to the force (\ref{her5}), 
\be
F\cong{1\over 4c}\left[-\alpha_Rn_{bg}+n_b\om\alpha_R'\right]{\pa\over\pa\tau}|\cE|^2.
\ee
Using (\ref{her5}) for this case, we obtain
\be
F\cong -{3\pi\epsoo a^3\over c}n_{bg}{n_s^2n_b^4\over(n_s^2+2n_b^2)^2}{\pa\over\pa\tau}|\cE|^2
\label{forceyyy}
\ee
if the dispersion of the dielectric material constituting the sphere is much smaller than that of the
host dielectric medium, i.e., if $d\eps_s/d\om\ll d\eps_b/d\om$. (Here $n_s$ is the refractive index at frequency $\om$ of the material of the sphere.) This result implies that, in the case of a slow-light host medium ($n_{bg}\gg 1$), the force on the sphere can be much larger than would be the case in a ``normally dispersive" medium, and is in the
direction opposite to that in which the field propagates.

The simple formula (\ref{forceyyy}), and similar expressions obtained in other limiting cases of (\ref{forceyy}),
obviously allow for a wide range of forces when a pulse of radiation is incident on a dielectric sphere in a host dielectric medium. Here we make only a few remarks concerning the last term in (\ref{forceyy}). Although
we have associated this contribution to the force with absorption, such a force appears even if the
sphere does not absorb any radiation of frequency $\om$. This is because there must be an imaginary
part of the polarizability simply because the sphere scatters radiation and thereby takes energy out of the
incident field. According to the optical theorem in this case of scattering by a nonabsorbing polarizable particle that
is small compared to the wavelength of the field, the imaginary part of the polarizability
is related to the complete (complex) polarizability as follows \cite{pwmetc}:
\be
\alpha_I(\om)={1\over 4\pi\epsoo}{2\om^3\over 3c^3}n_b|\alpha(\om)|^2.
\label{ot1}
\ee
Then the force proportional to $\alpha_I(\om)$ in (\ref{forceyy}) is
\be
F_{\rm scat}\equiv {1\over 2}n_b^5{\om\over c}\alpha_I|\cE|^2={8\pi\over 3}\left({\om\over c}\right)^4
{n_b^5I\over c}\left({\eps_s-\eps_b\over\eps_s+2\eps_b}\right)^2a^6,
\ee
which is just the well known ``scattering force" \cite{padgett} on a dielectric sphere in a medium with refractive index $n_b$, which may be taken to be real in the approximation in which the field is far from any absorption resonances
of the sphere.

\section{Conclusions}\label{sec:conclusions}
In this attempt to better understand the different electromagnetic momenta and the forces on electrically
polarizable particles in dispersive dielectric media, we have made several simplifications, including
the neglect of any surface effects, the treatment of the medium as a nondeformable body, and the approximation
of plane-wave fields. We have shown that conservation of momentum, even in seemingly simple examples
such as the Doppler effect, generally requires consideration not only of the
Abraham momentum and the Abraham force, but also of a contribution to the momentum of the medium due
specifically to the dispersive nature of the medium. We have generalized some well known expressions for the
forces on particles immersed in a dielectric medium to include dispersion. While we have presented arguments
in favor of the interpretation of the Abraham momentum as the momentum of the field, our simplified analyses
lead us to the conclusion that neither the Abraham nor the Minkowski expressions for momentum give the
recoil momentum of a particle in a dispersive dielectric medium. Finally we have shown that the force exerted
on a particle in a strongly dispersive medium is approximately proportional to the group index $n_{bg}$, and can therefore become very large in a slow-light medium.

\renewcommand{\theequation}{A-\arabic{equation}}
\setcounter{equation}{0}  
\section*{APPENDIX. Consistency of Eqs. (\ref{energydensity}) and (\ref{eqd}).}
Since the term involving $\alpha'$ in Eq. (\ref{eqd}) is essential to our analysis, and in particular to the derivation of Nelson's dispersive contribution [Eq. (\ref{nels})] to the momentum density, we review here the fact that the expression (\ref{energydensity}) for the total energy density may be regarded as a consequence of Eq. (\ref{eqd}) and Poynting's theorem. We begin by writing Poynting's theorem in its integral form:
\bea
\oint{\bf S}\cdot\hat{n}da&=&-\int\left[\bE\cdot{\pa{\bD}\over\pa t}+\mu_0\bH\cdot{\pa{\bH}\over\pa t}\right]dV
\nonumber \\
&=&-\int\left[{1\over 2}{\pa\over\pa t}(\epsoo\bE^2+\mu_0\bH^2)+\bE\cdot{\pa{\bf P}\over\pa t}\right]dV \nonumber \\
&=&-\int \dot{u}dV.
\label{loudon4}
\eea
The integral of the normal component of ${\bf S}=\bE\times\bH$ on the left-hand
side is, as usual, over a surface enclosing a volume $V$, and 
\be
\dot{u}={1\over 2}{\pa\over\pa t}[\epsoo\bE^2+\muo\bH^2]+\bE\cdot{\pa{\bf P}\over\pa t}.
\ee
$u$ is the density of total energy, that in the field plus that in the medium. Using ${\bf P}=N\bd$, together with
\be
\bE\cdot{\pa{\bf P}\over\pa t}={\pa\over\pa t}(\bE\cdot\bP)-{\pa{\bE}\over\pa t}\cdot{\bf P},
\ee
Eq. (\ref{eqd}), and $\bE=\cE_0\exp(-i\om t)$, we obtain
\be
\bE\cdot{\pa{\bf P}\over\pa t}={1\over 4}N(\alpha+\om\alpha'){\pa\over\pa t}|\cE_0|^2
\ee
when we take the cycle average and use the assumption made in obtaining (\ref{eqd}) that $\dot{\cE_0}$ is negligible compared to $\om\cE_0$. Then, from $\eps(\om)=\epsoo+N\alpha(\om)$, it follows that
\be
{1\over 2}{\pa\over\pa t}(\epsoo\bE^2)+\bE\cdot{\pa{\bf P}\over\pa t}={1\over 4}[\eps+\om{d\eps\over d\om}]
{\pa\over\pa t}|\cE_0|^2,
\ee
from which (\ref{energydensity}) follows.
\\ \\
{\bf Comments added after original submission:} \\ 

A dispersive contribution to the momentum such as appears in
Eq. (25), for example, appears also in earlier work by H. Washimi; see
H. Washimi and V.I. Karpman, Sov. Phys. JETP {\bf 44}, 528
(1976) and references therein. 

When the force (61) is applied to the case of a guest two-level atom
in a host medium it reduces to the expression given in Eq. (3) of
S.E. Harris, Phys. Rev. Lett. {\bf 85}, 4032 (2000) if it is
assumed that (i) the plane-wave field acting on the atom propagates
at the group velocity of the host medium; (ii) the field frequency 
is sufficiently different from the atom's transition frequency that 
the term proportional to $\alpha_R'$ is negligible; (iii) the medium
is sufficiently dispersive that $n_{bg}\gg n_b$; and (iv) absorption
is negligible, so that the term proportional to $\alpha_I$ may be
ignored.

We thank Dr. P.D. Lett for bringing the work of Washimi and Harris to
our attention.

\subsection*{Acknowledgement}
It is an honor to contribute to this special issue in memory of Krzysztof W\'odkiewicz, whose 
brilliance and passion for physics continue to inspire. DHB gratefully acknowledges support by the Los Alamos LDRD program. RWB and ZS gratefully acknowledge support by the 
NSF and by the DARPA/DSO Slow Light program. PWM acknowledges discussions relevant to this paper
with Paul R. Berman and, some years ago, with Nandor Balazs.


\begin{thebibliography}{99}
\bibitem{wod} Aside from a derivation of some results used in Section \ref{sec:sphere}, the material in this section, especially in connection with the Rayleigh attenuation coefficient, digresses somewhat from the main thrust of the paper. But because Krzysztof W\'odkiewicz was very much interested in historical matters of this sort, we feel it is not entirely inappropriate to include it.
\bibitem{rayleighnote} A proper derivation must, of course, include the particle density fluctuations,
but this leads to the same results and conclusions in the case of an ideal gas.
\end{thebibliography}
\end{document}